\documentclass[%
 aip,
 jcp,%
 amsmath,amssymb,
 reprint,%
citeautoscript,
]{revtex4-1}

\usepackage{graphicx}
\usepackage{dcolumn}
\usepackage{bm}
\usepackage{braket}
\usepackage{mathtools}
\usepackage[usenames,dvipsnames]{color}
\usepackage{natbib}

\begin{document}

\title{Mean Field Theory of Thermal Energy Transport in Molecular Junctions}

\author{Aaron Kelly}
\affiliation{Department of Chemistry, Dalhousie University, Halifax, Nova Scotia, Canada}

\date{\today}

\begin{abstract}
Mean field theory is applied to nonequilibrium thermal energy transport in a model molecular junction. An approximation to the total time-dependent heat current in the junction is constructed using an ensemble of Ehrenfest trajectories, and the average heat current in the steady state is obtained. The accuracy of this treatment is verified through benchmark comparisons with exact quantum mechanical results, and various approximate quantum transport theories, for the nonequilibrium spin-boson model. The performance of the multi-trajectory Ehrenfest approach is found to be quite robust, displaying good accuracy in intermediate cases that remain elusive to many perturbative approximations, and in the strong coupling limit where many methods break down. Thus, mean field theory and related trajectory-based approximate quantum dynamics methods emerge as a promising toolkit for the study of transport properties in nanoscale systems. 

\end{abstract}

%\keywords{nonequilibrium statistical mechanics, quantum dynamics, thermal energy transport}
\maketitle

\section{Introduction}
With rapid recent developments in molecular-scale electronics and a plethora of emerging nanotechnologies \cite{nat00,SCI07,natmat12,PRL14}, theoretical modelling of nonequilibrium quantum charge and thermal energy transport processes is becoming of increasing interest and importance \cite{nanodissipation2002,diventra2003,etransfer_tgradient_nitzan2016,carbogno_2017}. From the perspective of molecular simulation, progress in this area requires moving toward the treatment of real-time dynamics in realistic systems. This motivates the development of accurate and highly efficient methods to treat open quantum systems that can account for the effects of anharmonicity and nonadiabaticity beyond the perturbative and Markovian limits. 

In the cases where they can be converged, exact quantum mechanical solutions offer the benefit of unambiguous insight into the problem at hand. The current drawback is that range of systems for which exact solutions can be obtained remains rather narrow, and is essentially limited to generic systems with very few (less than twenty) degrees of freedom, or the class of quantum impurity models, which can have many degrees of freedom but a restricted functional form. Although the form of the impurity models may appear simple, they form the basis for simulations of a vast range of properties of condensed matter systems\cite{RMP96,RMP06}. Specifically, here we consider the nonequilibrium spin-boson model, which is a quantum impurity model that has become a workhorse for investigating the thermodynamics of open quantum systems\cite{RMP09,RMP11,rmp17} and nonequilibrium thermal energy transfer\cite{dvira05,dvira_annrev_2016}. As such, a number of exact solvers have been developed\cite{wangthoss08,saito_kondo_2013,chen17,makri17,gull18} that allow access to ever an increasing range of parameter space within this model. Nevertheless the need for approximate techniques that can go beyond this class of models is also clear. As such, these systems also serve to provide a rich testing ground for assessing the accuracy of approximate methods, in addition to providing physical insight into more realistic systems

The application of approximate quantum dynamics techniques to real-time nonequilibrium thermal transport was pioneered by Segal and Nitzan, beginning with their work on molecular junctions\cite{dvira_wires_2003,dvira_rectifier_2005,dvira_junction_2005}. This opened up a broad interest in this class of nonequilibrium transport problems, which has now become an active area of development and application for both exact quantum dynamics solvers\cite{wangthoss08,dvira_infpi_2013,saito_kondo_2013,kato2015,kato_2016,shi_heom_2017,dvira_ifcspi_2018} and approximate quantum transport theories including nonequilibrium Green's functions (NEGF) approaches\cite{mingo_NEGF_2006,NEGF_2006,NEGF_2007,NEGF_2011,wangthoss08,wangthoss10}, path integral techniques such as the noninteracting blip approximation (NIBA)\cite{dvira_niba_2011,dvira13} and other quantum master equations\cite{cao_polaron_qme_2015,junjie_niba_2017,dvira_majorana_2017}, as well as an approximation to the quantum-classical Liouville equation\cite{gabe18}. 

Here, mean field theory is employed to study thermal energy transport in a molecular junction, represented by the nonequilibrium spin-boson (NESB) model, in order to address the efficacy of quantum-classical trajectory-based dynamics methods for this problem. The remainder of this work is organized as follows: in Section II a brief overview of the NESB model is given, along with a description of the how mean field theory is employed to calculate time dependent heat currents. Simulation results for nonequilibrium steady state heat currents for various parameter regimes of the NESB model are reported in Section III, where a comparison with exact numerical benchmarks are reported where possible, as well with a range of other approximate quantum dynamics approaches. In Section IV an outlook to potential future work is offered, and some concluding remarks are given.   

\section{Theory}
\label{sec:theory}
\subsection{Nonequilibrium Spin-Boson Model}
The nonequilibrium spin - boson model model forms the basis of many recent studies of thermal energy transfer in nanoscale molecular junctions. Within this model, two vibrational states of the junction-bound molecule are considered, and each state is coupled to two phonon baths, labelled left (L) and right (R) respectively. The total Hamiltonian for the NESB system can be written in a standard system-bath form,
\begin{equation}
    \hat{H} = \hat{H}_s + \hat{H}_b + \hat{H}_{sb}.
    \label{eq:H_tot}
\end{equation} where $\hat{H}_s$ describes the two-level subsystem,
\begin{equation}
    \hat{H}_s = \frac{\epsilon}{2}\hat{\sigma}_z + \frac{\Delta}{2}\hat{\sigma}_x,
    \label{eq:FE_H_s}
\end{equation} and $\epsilon$ and $\Delta$ are related to the coupling between the two vibrational states of the molecule, and their energy difference, respectively. 

The Hamiltonian operator for the environment includes both the left and right baths, written using mass-weighted coordinate and momenta operators is 
\begin{equation}
    \hat{H}_b = \frac{1}{2} \sum_{\lambda=L,R}\sum_{k}\Big[\hat{P}_{\lambda,k}^2 +  \omega^2_{\lambda,k}\hat{Q}^2_{\lambda,k}\Big].
    \label{eq:FE_H_b}
\end{equation}

\begin{figure}
 \includegraphics[width=\columnwidth]{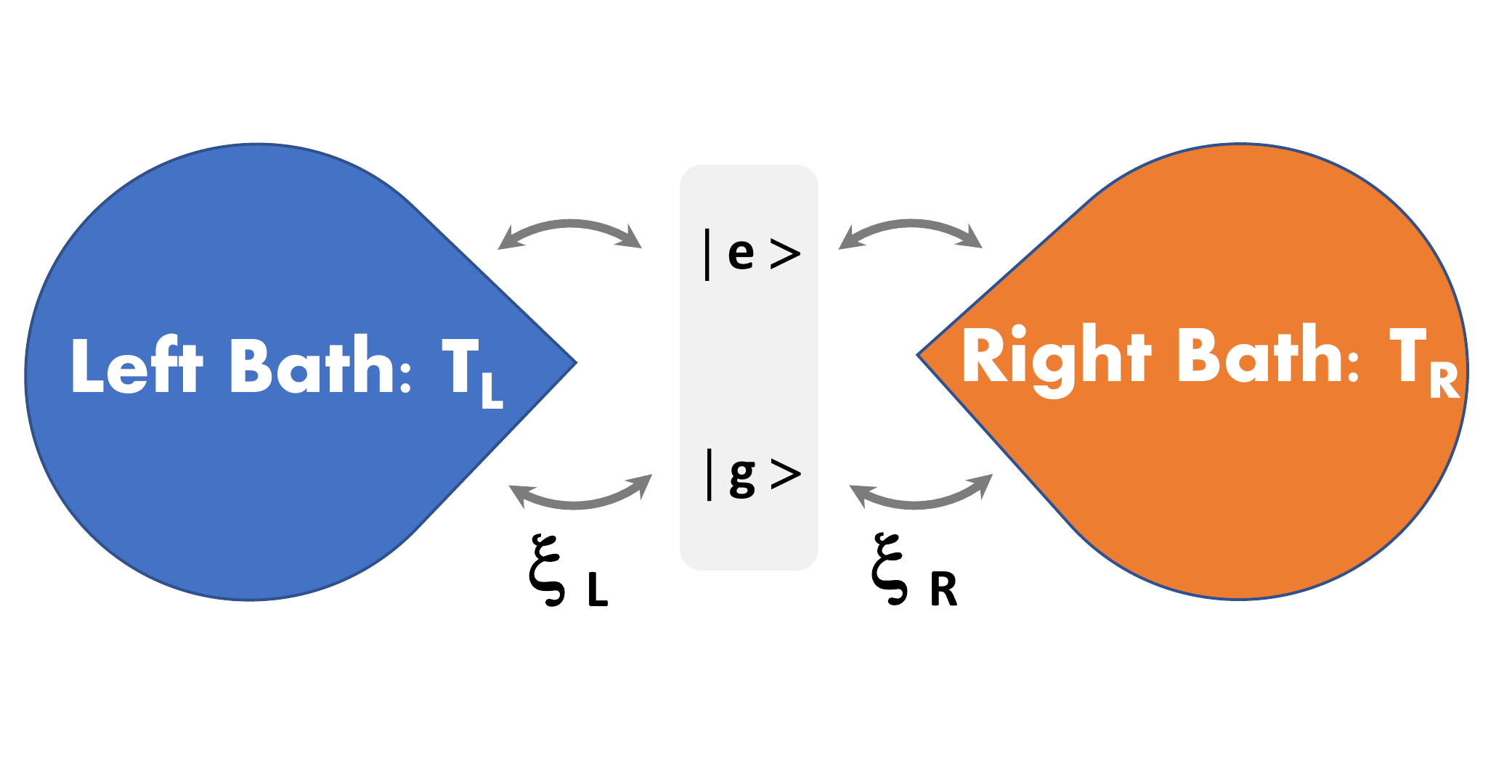}
  \caption{Cartoon of the nonequilibrium spin-boson model. The ground and excited states of the junction-bound molecule are labelled $|g\rangle$ and $|e\rangle$, and are coupled to both the left and right thermal baths, with coupling strengths $\xi_L$ and $\xi_R$ respectively.}
\label{fig1}
\end{figure}

The molecular subsystem is bilinearly coupled to each bath, 
\begin{equation}\label{eq:FE_H_sb}
\begin{split}
    \hat{H}_{sb} &= \sum_{\lambda=L,R}\sum_{k} c_{\lambda,k} \hat{Q}_{\lambda,k} \hat{\sigma}_z.
\end{split}
\end{equation} This coupling is characterized by a spectral density function,
\begin{equation}
    g_{\lambda}(\omega) = \frac{\pi}{2} \sum_{k} \frac{c_{\lambda,k}^2}{\omega_{\lambda,k}} \delta(\omega - \omega_{\lambda,k}),
\end{equation}
and, as is common in the literature\cite{dvira05,dvira_niba_2011,wangthoss08,wangthoss10}, we choose the Ohmic form in this study.
\begin{equation}
    g_{\lambda}(\omega) = \frac{\pi}{2}\xi_{\lambda} \omega\exp(-\omega/\omega_{c,\lambda}).
\end{equation} Here the Kondo parameter, $\xi_{\lambda}$, is a measure of the system-bath coupling strength at the $\lambda^{th}$ contact, and $\omega_{c,\lambda}$ is the inverse of the corresponding characteristic timescale of each bath. In this study the right and left bath spectral densities are chosen to be identical, i.e. $\xi_L = \xi_R= \xi$, and $\omega_{c,L} = \omega_{c,R}= \omega_c$, such that the baths only differ in terms of their temperature.

\subsection{Mean Field Theory}
The multi-trajectory Ehrenfest mean field theory (MFT) can be derived in a straightforward manner, via the quantum-classical Liouville (QCL) equation\cite{qcle,Grunwald2009}. The QCL equation of motion for the density matrix is formally exact for an arbitrary quantum mechanical system that is bilinearly coupled to a harmonic environment, as is the case for the NESB problem, and many other quantum impurity models. Written in a compact form, the QCL equation is \begin{eqnarray} \label{eq:qcle}
  && \frac{\partial }{\partial t}\hat{\rho}_W (X, t) =-i {\mathcal L}\hat{\rho}_W (X,t).
\end{eqnarray} The QCLE describes the time evolution of $\hat{\rho}_W(X,t)$, the partial Wigner transform of the density operator taken over the coordinates of the $N_b$ bath degrees of freedom, which are represented by continuous phase space variables $X=(R,P)=(R_1,R_2,...,R_{N_b},P_1,P_2,...,P_{N_b})$.  The partial Wigner transform of the density operator, $\hat{\rho}$, is defined as
\begin{eqnarray} \label{eq:wigner}
\hat{\rho}_{W}(R,P) = \frac{1}{(2\pi\hbar)^{N_b}}\int dZ e^{i P \cdot Z} \langle R - \frac{Z}{2} | \hat{\rho} | R +\frac{Z}{2}\rangle.  
\end{eqnarray} The QCL operator is defined as
\begin{equation}\label{eq:qcl_op}
i{\mathcal L} \cdot = \frac{i}{\hbar}[\hat{H}_W,\cdot] - \frac{1}{2}(\{\hat{H}_W,\cdot\}
-\{\cdot,\hat{H}_W\}),
\end{equation}
where $[\cdot,\cdot]$ is the commutator, and $\{\cdot,\cdot\}$ is the Poisson bracket in the phase space of the environmental variables. The Ehrenfest MFT equations of motion are obtained by requiring that the total density of the system can be written as an uncorrelated product of the system and bath reduced densities at all times,
\begin{equation}
\hat{\rho}_W(X,t)=\hat{\rho}_s(t) \rho_{b,W}(X,t),
\end{equation} 
where the reduced density matrix of the system is
\begin{equation}\label{eq:rdm}
\hat{\rho}_{s} (t) = Tr_b \Big( \hat{\rho}(t) \Big) = \int dX \hat{\rho}_W (X,t),
\end{equation} and the bath density is $\rho_b(X, t) = Tr_s ( \hat{\rho}_W (X,t))$. 
Requiring solutions to the QCL equation of this form yields the Ehrenfest MFT equations of motion for the subsystem:
\begin{equation}
    \frac{d}{dt} \hat{\rho}_s(t) = -i\Big[ \hat{H}_s + \hat{H}_{sb,W}(X(t)), \hat{\rho}_s(t)\Big].
\end{equation}
The evolution of the reduced Wigner density of the bath can be represented by an ensemble of multiple independent trajectories, $\rho_{b,W} (X,t) = \sum_j \delta(X_j-X(t))$, that evolve according to a set effective Hamilton's equations of motion generated from the mean-field effective Hamiltonian,
\begin{equation}
    \frac{\partial R_{\alpha}}{\partial t} =  \frac{\partial H_{b,W}^{Eff}}{\partial P_{\alpha}},\quad \frac{\partial P_{\alpha}}{\partial t} = - \frac{\partial H_{b,W}^{Eff}}{\partial R_{\alpha}}.
\end{equation} The effective Hamiltonian in the classical equations of motion is, \begin{equation}
    H^{Eff}_{b,W} = H_{b,W} + Tr_s\Big(\hat{H}_{sb,W}(X,t)\hat{\rho}_s(X,t)\Big) .
\end{equation}
The exact expression for the average value of any observable, $\langle O (t) \rangle $, can be written as \begin{eqnarray} 
\langle O (t) \rangle  = Tr_s \int dX  \hat{O}_W(X,t) \hat{\rho}_W(X,0).
\end{eqnarray} The mean field limit of this expression simple corresponds to evaluating the integral by sampling initial conditions for an ensemble of independent trajectories from $\hat{\rho}_W(X,0)$, and then generating the time evolution for each trajectory by approximating $\hat{O}_W(X,t)$ by it's mean-field counterpart.   

\subsection{Thermal Energy Transport: Observables of Interest}
The transport of thermal energy in the NESB model can be monitored via time-dependence of the average energy of each bath, \begin{equation}
    \langle H_{b,\lambda}(t)\rangle  = Tr \Big( \hat{H}_{b,\lambda}(t) \hat{\rho}(0) \Big).
\end{equation} 
Initially, we imagine that the system is separable; each bath is assumed to be in a canonical equilibrium state with temperature $T_{\lambda}$, and the two-level system is initially in the excited state. This corresponds to the following density operator, \begin{equation}
\hat{\rho}(0) = \hat{\rho}_{b,L}^{eq} \otimes \hat{\rho}_{b,R}^{eq} \otimes \hat{P}_s^{22},  
\end{equation} where $\hat{P}_s^{\alpha \alpha'} = | \alpha \rangle \langle \alpha' |$ is a subsystem projector. The bath initial conditions are sampled from the Wigner transform of the initial canonical density operators, with inverse temperature $\beta_{\lambda} = (k_BT_{\lambda})^{-1}$,   
\begin{equation}
\begin{split}
    (\rho_{b,\lambda}^{eq})_W &= \prod_{k}\frac{\tanh(\beta_\lambda \omega_{\lambda,k}/2)}{\pi} \\
    &\ \times \exp\Bigg[ - \frac{\tanh(\beta_\lambda \omega_{\lambda,k}/2)}{\omega_{\lambda,k}}\Big[ P_{\lambda,k}^2 + \omega_{\lambda,k}^2 R_{\lambda,k}^2\Big]\Bigg],
\end{split}
\end{equation} 

The nonequilibrium heat currents flowing through the junction are then defined as the time rate of change of the respective average bath energies, 
\begin{eqnarray}
    J_{L} &=& \lim_{t \rightarrow \infty} J_{L} (t) = \lim_{t \rightarrow \infty} \frac{d}{dt} \langle H_{b,L} (t) \rangle, \\
    J_{R} &=& \lim_{t \rightarrow \infty} J_{R} (t) = \lim_{t \rightarrow \infty} -\frac{d}{dt} \langle H_{b,R} (t) \rangle, 
\end{eqnarray} and the total nonequilibrium steady state (NESS) heat current is then defined as the long time limit of the total heat current through the junction, 
\begin{equation}
    J = \lim_{t \rightarrow \infty} J (t) = \lim_{t \rightarrow \infty} \frac{1}{2} \Big( J_{L} (t) + J_{R} (t) \Big). 
\end{equation}

\section{Results and Discussion}
\begin{figure}
  \includegraphics[width=\columnwidth]{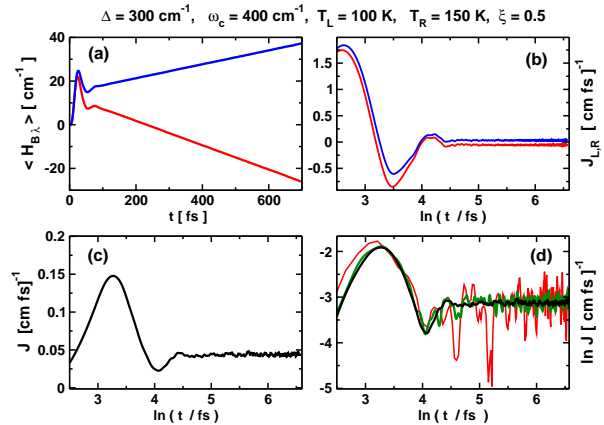} 
  \caption{Time evolution of (a) the bath energies $\langle H_{b,\lambda}(t)\rangle$, blue is the cold bath (L) and red is the warm bath (R), (b) the components of the total heat current, $J_{L,R}(t)$ versus log-scale time, blue (L) and red (R)(c) the total heat current, $J(t)$ versus log-scale time, and (d) log-log plot of $J(t)$ showing the convergence properties using three different trajectory ensemble sizes: $N_{traj} = 10^5$ (red), $N_{traj} = 10^6$ (green), and $N_{traj} = 10^7$ (black).} 
  \label{2}
\end{figure}
The results of multi-trajectory Ehrenfest mean field theory simulations of the NESS thermal energy transfer rate in the NESB model, with $\epsilon=0$, are summarized. In all simulations, at $t=0$ the two uncoupled reservoirs are then brought into contact with the subsystem. After a short period of transient dynamics, thermal energy steadily flows through the junction along the direction of the thermal gradient, as shown in Fig. 2. The steady-state heat current is reached as the rate of the energy flux between the left and right baths reaches a constant value, which is depicted in panels (b - d) of Figure 2. Somewhat surprisingly, the statistical properties of the thermal energy transfer dynamics are different than those of the population dynamics of the junction-bound molecular system. For example, the MFT simulations are found to be quite well converged for molecular properties using  $\sim10^4$ trajectories, while the corresponding heat current calculation requires two to three orders of magnitude more trajectories for convergence. 

\begin{figure}
  \includegraphics[width=\columnwidth]{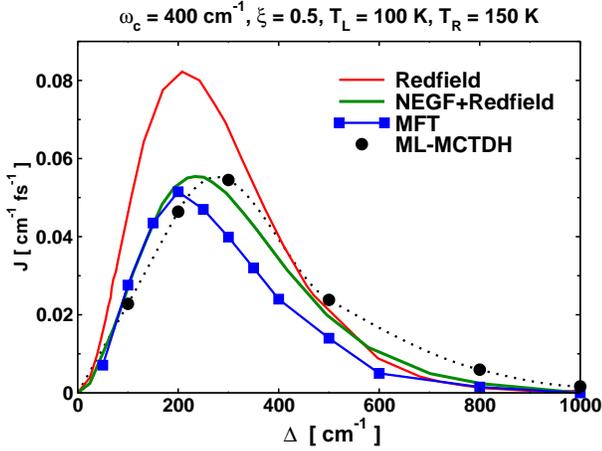}
  \caption{NESS heat current versus the vibrational splitting of the two level system, $\Delta$. Exact results are taken from Ref. 16.}
\label{3}
\end{figure}
\begin{figure}
 \includegraphics[width=\columnwidth]{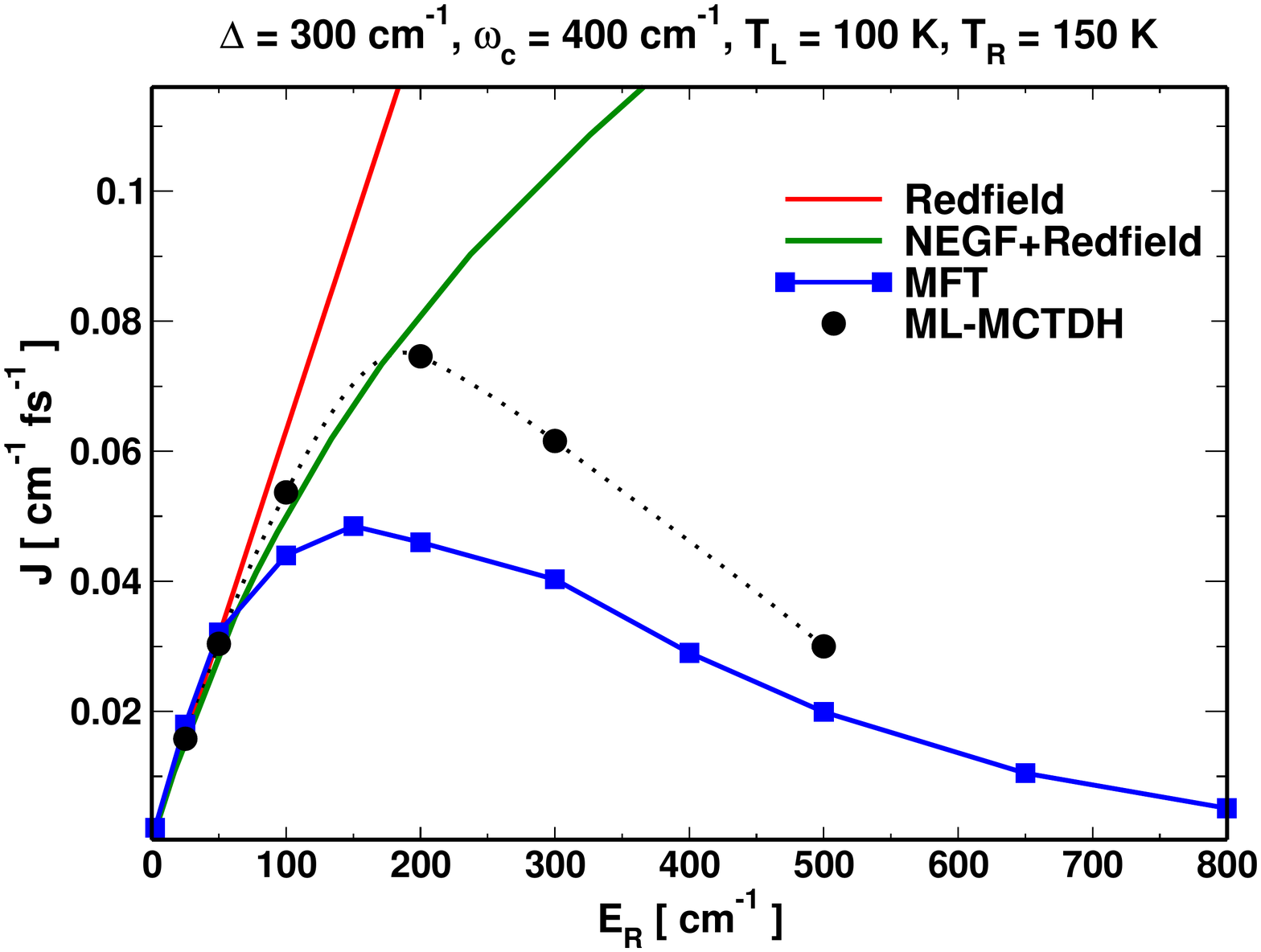}
  \caption{NESS heat current versus the total reorganization energy, for $\omega_c/\Delta = 4/3$. Exact and NEGF results are taken from Ref. 33.}
\label{4}
\end{figure}

\begin{figure}
   \includegraphics[width=\columnwidth]{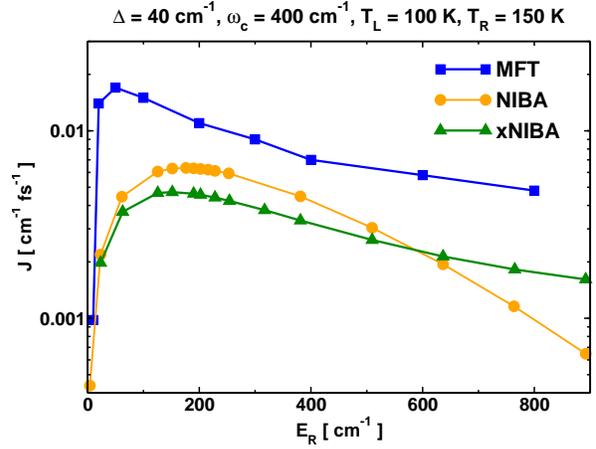} 
  \caption{NESS heat current versus the reorganization energy, for $\omega_c/\Delta = 10$. NIBA and xNIBA results are taken from Ref. 34.}
\label{5}
\end{figure}

\begin{figure}
 \includegraphics[width=\columnwidth]{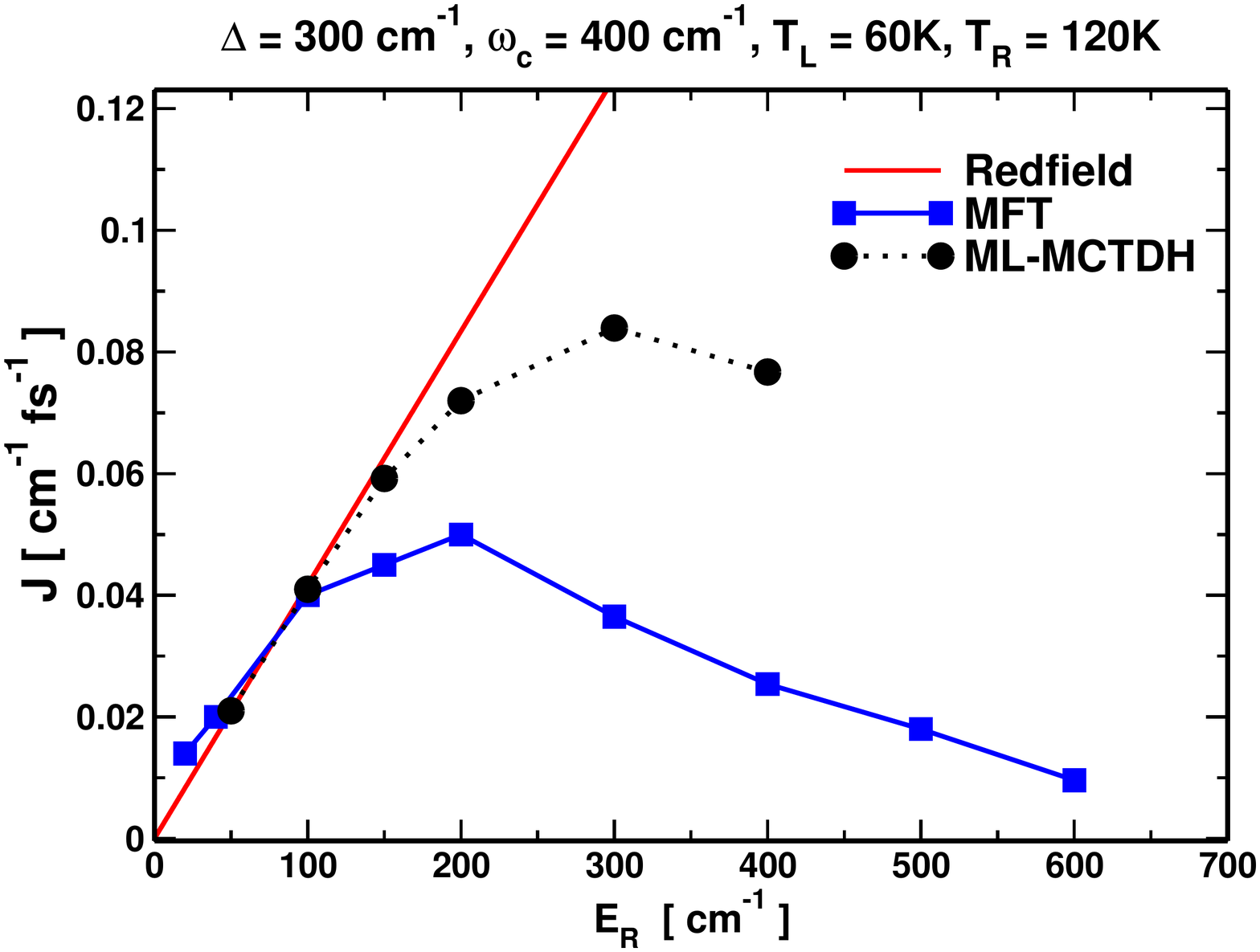}
  \caption{NESS heat current versus the reorganization energy, for $\omega_c/\Delta = 4/3$. Exact and NEGF results are taken from Ref. 33.}
\label{6}
\end{figure}

\begin{figure}
 \includegraphics[width=\columnwidth]{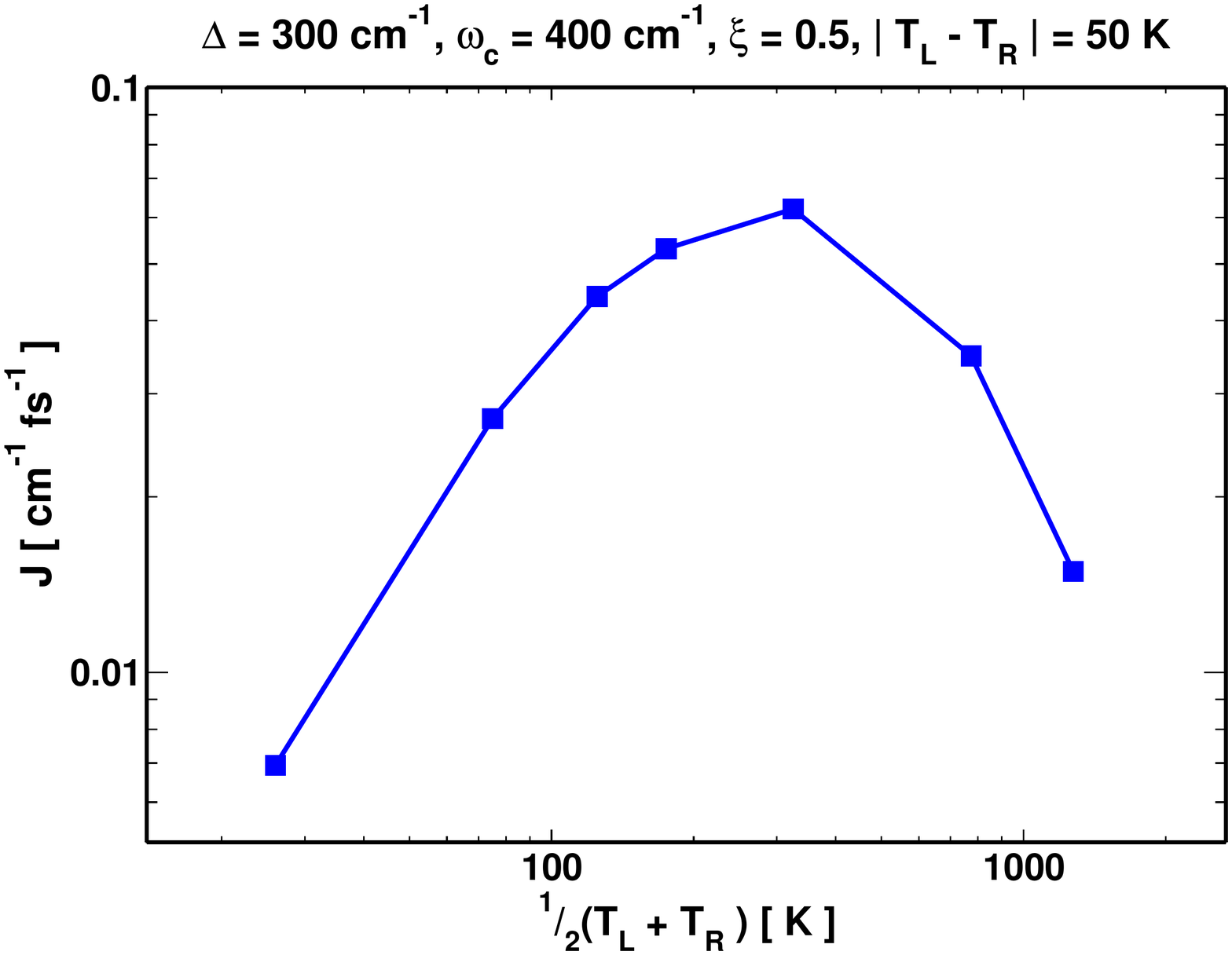}
  \caption{Log-log plot of the mean field NESS heat current versus the average bath temperature, at a constant temperature difference of $50$K between the left and right baths.}
\label{7}
\end{figure}

The dependence of the NESS heat current on the vibrational splitting of the molecular system, $\Delta$, is depicted in Figure 3. In this case, the MFT results are compared with numerically exact multi-layer multi-configurational time-dependent Hartree (ML-MCTDH) benchmark calculations\cite{wangthoss08}, as well as Redfield theory\cite{dvira05}, and a hybrid NEGF-Redfield approach\cite{wangthoss10}. All of these approaches capture the qualitative turnover behaviour of the current quite well in this case, with the MFT and NEGF approaches showing almost quantitative agreement across the full parameter range. 

Next, in Fig. 4, the NESS heat current is shown as a function of the system-bath coupling strength, which is related to the vibrational reorganization energy of the molecular junction, $E_R = \frac{\omega_c \xi}{2}$. Both the Redfield and NEGF-based theories fail to capture the correct turnover behaviour for the heat current, while the MFT results nicely reproduce the exact trend. The MFT result consistently underestimates the total heat current through the intermediate coupling regimes, but the qualitative agreement with the ML-MCTDH benchmark from weak to strong coupling is impressive, and overall is a significant improvement over the other approaches. 

A strongly nonadiabatic regime is depicted in Figure 5, with $\omega_c/\Delta = 10$. In this regime the MFT approach is compared with the non-interacting blip approximation (NIBA) and it's extension (xNIBA), which is expected to be more accurate\cite{dvira_niba_2011}. Again we see very good agreement between the theories, across the entire parameter regime, with MFT and xNIBA showing a very similar shape, especially in the strong coupling regime. 

In Fig. 6, the NESS heat current is investigated in another rather challenging scenario; a low-temperature system with a sizeable temperature gradient, and strong vibrational coupling, tuning between weak to strong system-bath coupling regimes. Overall, we see that the MFT approach emerges as a rather accurate approximate approach. Figure 7 shows the behaviour of the heat current through the junction as a function of the average temperature of the the two baths, at a fixed temperature bias. The turnover displayed here mirrors the trends shown in the other parameters, and is representative of the underlying resonant nature of the nonequilibrium transport process.

\section{Conclusions and Outlook}
Obviously not a panacea, this version of the mean field approach is expected to break down in some cases, for example at very low temperatures, or for very fast or strongly correlated baths. Furthermore, when $\epsilon$ becomes nonzero, MFT may also suffer from the breakdown of detailed balance, analogous to it's performance for subsystem properties in the traditional, single bath, spin-boson model. In this respect, more accurate trajectory-based techniques based on mean field theory\cite{ctmqc,sato_2018}, or others stemming from the quantum-classical Liouville equation\cite{qcle}, or the partially linearized path integral formalism\cite{coker08}, may prove useful.  

One technical item of note is that relatively large trajectory ensembles were needed to converge the average heat current, as compared to subsystem properties. Indeed, recent MFT studies of spontaneous emission\cite{norah_2019} show very similar convergence behaviour for subsystem versus bath properties to that observed here. Potentially, other forms for the current operator could be considered, which may have different statistical convergence properties. For example, adopting a form which jointly depends on the junction and bath degrees of freedom may be useful in this respect. To further improve the efficiency, and potentially the accuracy, of this description the mean-field generalized quantum master equation approach, which has previously proven successful in treating the dynamics of systems coupled to harmonic \cite{kelly15,andres16,kelly16,andres17} or fully atomistic environments\cite{jpcl15}, could be explored in this more general setting using an extensions of the Shi-Geva formalism\cite{shi_geva_2003,geva2006,geva2019} for non-system operators\cite{cohen_nonsys_2013}. 

The multi-trajectory implementation of Ehrenfest mean field theory thus presents surprisingly accurate, and hence rather promising, tool for the description of nonequilibrium thermal energy transport in molecular junctions. This approach lays the foundation for future trajectory-based studies of nonequilibrium steady state transport, such as current fluctuations, higher order counting statistics, and thermal conductivity, in both impurity models for molecular junctions and more realistic systems. 

\acknowledgments
AK acknowledges financial support from the National Sciences and Engineering Research Council (NSERC) of Canada through the Discovery Grant program, and would like to thank Dvira Segal and Gabriel Hanna for helpful comments and insightful discussions.

\bibliography{bibliography}

\end{document}